# Deep Convolutional Neural Network Model for Short-Term Electricity Price Forecasting

Hsu-Yung Cheng
*Department of Computer Science and Information Engineering, National Central University*

Ping-Huan Kuo
*Computer and Intelligent Robot Program for Bachelor Degree, National Pingtung University*

Yamin Shen and Chiou-Jye Huang*
*School of Electrical Engineering and Automation, Jiangxi University of Science and Technology*

Corresponding author. Tel.: +86 18070585481

E-mail: n28921094@gs.ncku.edu.tw

ABSTRACT: In the modern power market, electricity trading is an extremely competitive industry. More accurate price forecast is crucial to help electricity producers and traders make better decisions. In this paper, a novel method of convolutional neural network (CNN) is proposed to rapidly provide hourly forecasting in the energy market. To improve prediction accuracy, we divide the annual electricity price data into four categories by seasons and conduct training and forecasting for each category respectively. By comparing the proposed method with other existing methods, we find that the proposed model has achieved outstanding results, the mean absolute percentage error (MAPE) and root mean square error (RMSE) for each category are about 5.5% and 3, respectively.

INTRODUCTION

With the development of energy market, the electricity industry is becoming more market-oriented. The precise price trend forecasting can help producers and traders make better decisions and avoid potential risk.

Electricity price fluctuation is largely determined by electricity demand fluctuation. electricity demand prediction can be affected by multiple factors, such as season, holiday, temperature etc. Many methods of electricity price and demand forecasting [1-3] have been applied.

Chen et al. [4] proposed BP neural networks based on genetic simulated annealing algorithm (GSAA) to forecast the short-term electricity price, the author made a comparation between the BP and GSAA-BP, and the results show that GSAA-BP have a good performance in electricity price prediction. Chang et al. [5] applied the Adam-optimized long short-term memory network (LSTM) to forecast the price of electricity which is a type of recurrent neural network (RNN) used in deep learning. Kuo et al. [6] combined LSTM with CNN to predict the electricity price. Zhou et al. [7] proposed an optimized complex structure LSTM to predict the electricity price, they applied the ensemble empirical mode decomposition (EEMD) to process the data and used the bayesian optimization to select the hyperparameters.

The main framework of this paper is as follows. The method is described in section II. section III introduces the data selection and evaluates the results of the proposed model. Finally, the conclusion is summarized in section VI.

## METHODOLOGY

The architecture of the proposed model as show in figure 1, It consist of three convolution layers and three maxpooling layers, the activation functions totally adopt the ReLU，as show in (1)

$$\text{ReLU}(x) = \max(0, x) \qquad (1)$$

The input of proposed model is the electricity price in the past 23 h, and the output is the electricity price in the next hour. Convolution layers and pooling layers are used to extract features, fully connected layers are used to fit the unknown objective function through the features which obtained by Convolution layers and pooling layers

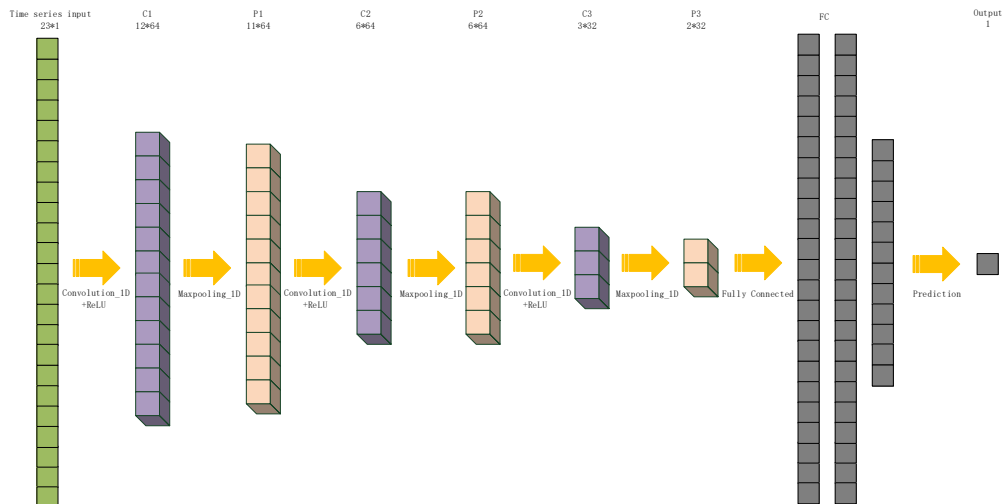

Figure 1. The architecture of the proposed model.

## EXPERIMENTS

A. Data Selection & Analysis

The total data set consists of hourly electricity pricing information for New York City from 2015 to 2017 provided by ENGIE resources. The time-series data of electricity price for the full year of 2017 is show in figure 2.

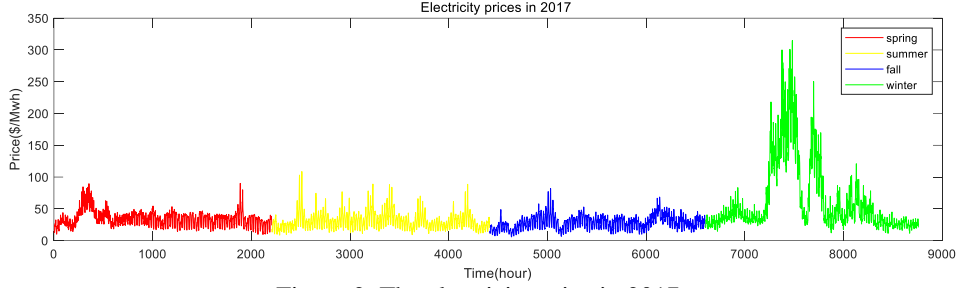

Figure 2. The electricity price in 2017

In this paper, we use the electricity price from 2015 to 2017 as the training set and 2018 as the test set.

New York is in the northern hemisphere of the earth. March to May is spring, June to August is summer, September to November is fall and December to next year February is winter. Therefore, we combine the four seasons which are mentioned before into one year. As show in figure 2. It is obviously show that the fluctuation trend of electricity price in the same season is similar, but it is different in different seasons, especially in winter. Hence, in order to avoid the influence of different seasonal fluctuations on model prediction accuracy, we divide the annual electricity price data into four categories by season and conduct training and prediction for each category, respectively.

B. Evaluation Criteria

There are two statistical methods, the mean absolute percentage error (MAPE) and root mean square error (RMSE),were applied to evaluate the performance of forecasting model. The formulas can be expression as follows:

$$\text{MAPE} = \frac{1}{N}\sum_{i=1}^{N}\frac{|y_i - \hat{y}_i|}{y_i} \times 100\% \quad (2)$$

$$\text{RMSE} = \sqrt{\frac{1}{N}\sum_{i=1}^{N}(y_i - \hat{y}_i)^2} \quad (3)$$

N—total number of prediction points
$y_i$—the actual electricity price value of the $i^{th}$ prediction point
$\hat{y}_i$— the predicted electricity price value of the $i^{th}$ prediction point

C. Experimental Results

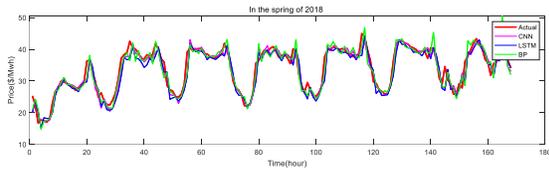

(a) The comparisons of all forecasting results in spring

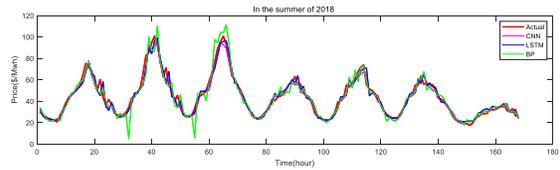

(b) The comparisons of all forecasting results in summer

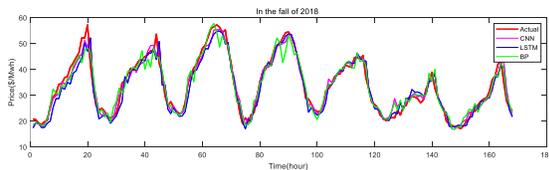

*(c) The comparisons of all forecasting results in fall*

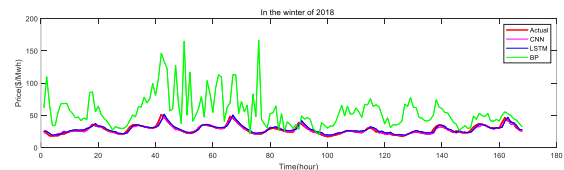

(d) The comparisons of all forecasting results in winter

Figure 3. The comparisons of all forecasting results in different seasons

Figure 3 shows the comparisons of all forecasting results in different seasons, in the figure 3, we have selected actual and forecast electricity prices for the first week of April, July, October in 2018 and the first week of January in 2019 to display the results. MAPE and RMSE for different models in different seasons are show in table 1 and table 2. It can be seen that both the CNN and LSTM show the good performance in any season, especially CNN has the outstanding performance in winter, BP has the worst performance. In general, the proposed model shows the best robustness and accuracy, compared with LSTM and BP.

TABLE I
THE MAPE OF ALL FORECASTING RESULTS IN DIFFERENT SEASONS

| season | BP | LSTM | CNN |
|---|---|---|---|
| Spring | 5.59 | 5.83 | 5.36 |
| Summer | 6.03 | 5.63 | 5.18 |
| Fall | 7.35 | 5.95 | 5.38 |
| Winter | 11.58 | 6.65 | 5.49 |
| Average | 7.64 | 6.02 | 5.35 |

TABLE II
THE RMSE OF ALL FORECASTING RESULTS IN DIFFERENT SEASONS

| season | BP | LSTM | CNN |
|---|---|---|---|
| spring | 2.30 | 2.44 | 2.19 |
| summer | 3.83 | 3.19 | 2.85 |
| fall | 5.96 | 3.36 | 3.03 |
| winter | 22.66 | 4.70 | 4.13 |
| Average | 8.69 | 3.42 | 3.05 |

CONCLUSIONS

In this paper, CNN is applied to predict the hourly electricity price and each season hourly electricity price forecasting results were exhibited. Simulation results demonstrate that CNN has outstanding performance than LSTM and BP.

This study has two shortcomings, firstly, the data analysis of the price sequence without considering the impact of holidays, but the price anomalies caused by holidays will affect the accuracy of the prediction. Secondly, the selection of hyperparameters is made through multiple experiments, which may lead to the fact that the hyperparameters we choose may not be the optimal. Therefore, in the future experiments, we can optimize the model by improving the above shortcomings